\documentclass[aps,prl,twocolumn,amsmath,amssymb]{revtex4-2}
\usepackage{amsmath,amsfonts,amssymb,bm,hyperref,xcolor,amsopn}
\usepackage[capitalize]{cleveref}
\usepackage{graphicx,subfigure,epstopdf,multirow,diagbox,array}
\usepackage[normalem]{ulem}
\usepackage{soul}
\usepackage{cancel}
\usepackage{mathrsfs}

\soulregister\cite7
\soulregister\ref7

\begin{document}

\title{Hierarchical and ultrametric barriers in the energy landscape of jammed granular matter}

\author{Shuonan Wu}
\thanks{These two authors contributed equally}
\affiliation{School of Mathematical Sciences, Peking University, Beijing 100871, China}

\author{Yuchen  Xie}
\thanks{These two authors contributed equally}
\affiliation{School of Mathematical Sciences, Peking University, Beijing 100871, China}

\author{Deng Pan}
\email{dengpan@mail.itp.ac.cn}
\affiliation{Institute of Theoretical Physics, Chinese Academy of Sciences, Beijing 100190, China}

\author{Lei Zhang}
\email{zhangl@math.pku.edu.cn}
\affiliation{Beijing International Center for Mathematical Research, Peking University, Beijing, 100871, China}
\affiliation{Center for Quantitative Biology, Peking University, Beijing, 100871, China}
\affiliation{Center for Machine Learning Research, Peking University, Beijing, 100871, China}

\author{Yuliang Jin}
\email{yuliangjin@mail.itp.ac.cn}
\affiliation{Institute of Theoretical Physics, Chinese Academy of Sciences, Beijing 100190, China}
\affiliation{School of Physical Sciences, University of Chinese Academy of Sciences, Beijing 100049, China}
\affiliation{Center for Theoretical Interdisciplinary Sciences, Wenzhou Institute, University of Chinese Academy of Sciences, Wenzhou, Zhejiang 325001, China}

\begin{abstract}
According to the mean-field glass theory, the (free) energy landscape of disordered systems is hierarchical and ultrametric if they belong to the full-replica-symmetry-breaking  universality class. However, examining this theoretical picture in three-dimensional systems remains challenging, where the energy barriers become finite. Here, we numerically explore the energy landscape of granular models near the jamming transition using a saddle dynamics algorithm to locate both local energy minima and saddles. The multi-scale distances and energy barriers between minima are characterized by two metrics, both of which exhibit signatures of an ultrametric space. The scale-free distribution of energy barriers reveals that the landscape is hierarchical.
\end{abstract}
\maketitle

\begin{figure}
    \centering
    \includegraphics[width=\linewidth]{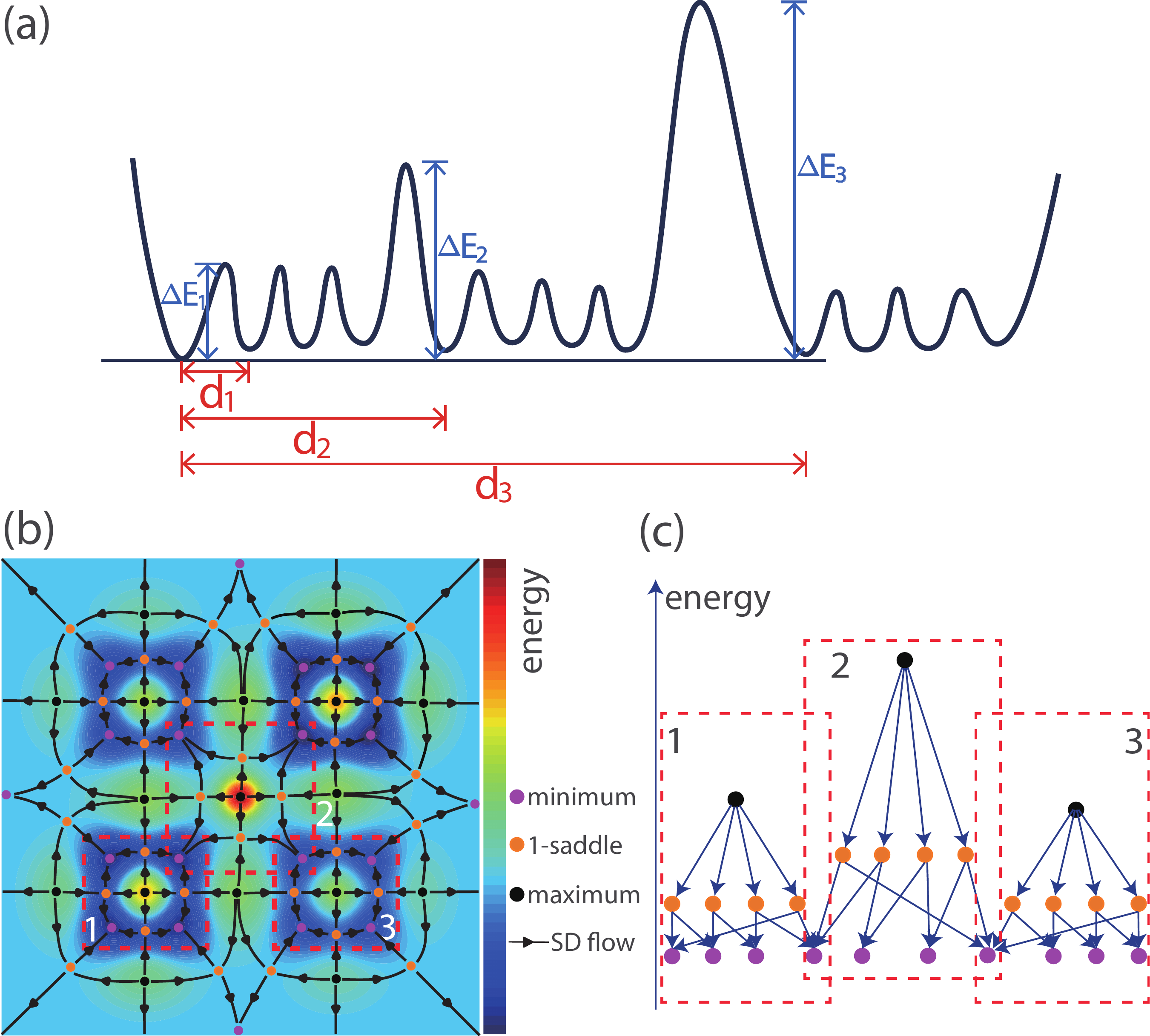}
    \caption{{\bf Energy landscape and pathway map.}
    (a) A schematic multi-scale energy landscape. The inter-minima distance $d$ and barrier height $\Delta E$ both indicate the multi-scale nature of the landscape. 
    (b) The pathway map of a two-scale, two-dimensional energy surface $E(x,y)$.  {The expression of $E(x,y)$ is shown in Supplemental Material (SM) Sec.~S1.}
     (c) Tree representation of the pathway map. We only show the stationary points in the three red dashed boxes.
    }
    \label{fig1}
\end{figure}

{\bf{Introduction.}}
The complex (free) energy landscapes of amorphous materials govern many of their physical properties, including activated dynamics and mechanical response to deformation~\cite{debenedetti2001theory, charbonneau2014fractal, jin2017exploring}. These landscapes are described by an energy function \( E(\bf{x}) \) of \( N \times d_{\mathrm{f}} \)-dimensional coordinates \( \bf{x} \), where \( N \) is the number of particles and \( d_{\mathrm{f}} \) the degrees of freedom per particle. Typically, the landscape is characterized by numerous local minima corresponding to metastable states, and saddle points with at least one unstable direction. Understanding the organization of these minima and saddles is essential and challenging. A useful approach involves characterizing the statistical properties of the multi-scale distance \( d \) between minima, the multi-scale barrier height \( \Delta E \) for minimum-saddle-minimum activated pathways, and the relationship between them (Fig.~\ref{fig1}a).

On the theoretical side, the replica theory has yielded exact solutions for several mean-field glass models, including the Sherrington--Kirkpatrick spin glass~\cite{mezard1987spin}, and hard/soft spheres in infinite dimensions~\cite{charbonneau2014fractal, charbonneau2014exact, biroli2016breakdown, biroli2018liu, Parisi_Urbani_Zamponi_2020}. The energy landscapes of these models are universally described by a full-step replica symmetry breaking (full-RSB) solution, which reveals that the distance metric \(d\) between local minima is \emph{hierarchical and ultrametric}. In amorphous solids, such a landscape emerges in the so-called \emph{Gardner phase} near the jamming transition~\cite{gardner1985spin, berthier2019gardner, urbani2023gardner, berthier2016growing, wang2024gardner}.

More recently, theoretical efforts have extended to analyzing saddle points in mean-field spin glass models~\cite{ros2019complexity, ros2021dynamical, kent2023count, kent2024arrangement}. Furthermore, mean-field theories have been used to study the power-law avalanche distribution \(f(\Delta E) \sim \Delta E^{-\tau}\) in response to an external field, where $\Delta E$ is the energy drop associated to barrier crossing.
Models of the critical flow state predict \(\tau_{\rm MF} = 3/2\)~\cite{sethna1993hysteresis, fisher1997statistics, fisher1998collective, dahmen2009micromechanical, dahmen2011simple, otsuki2014avalanche}, while  the full-RSB theory suggests \(\tau_{\rm J} \approx 1.41269\) at the jamming transition ($\varphi = \varphi_{\rm J}$) and \(\tau_{\rm UNSAT} = 1\) above it ($\varphi > \varphi_{\rm J}$)~\cite{franz2017mean}, where $\varphi_{\rm J}$ is the jamming transition density. Despite these advances, the properties of saddle points remain less understood than those of minima. In particular, it is unclear whether saddles are also organized hierarchically and ultrametrically, as has been established for minima.

In two and three dimensions (2D and 3D), theoretical analyses are intractable, so exploring the energy landscape must rely on numerical simulations. The hierarchical structure of the energy landscape in granular matter near the jamming transition has been evidenced by the measured ultrametric distances ($d$) between minima~\cite{liao2019hierarchical, dennis2020jamming, artiaco2020exploratory}. However, studies on saddle points and the associated energy barriers are limited~\cite{keyes1997instantaneous, gezelter1997can, angelani2000saddles, broderix2000energy, doye2002saddle, denny2003trap}. 
Due to the difficulty in the direct measurement of energy barriers, 
their distributions $f(\Delta E)$ are often inferred indirectly from accessible physical processes. For instance, in athermal systems such as driven granular matter, a power-law distribution of energy drops, \(f(\Delta E) \sim \Delta E^{-\tau}\), is universally observed during plastic events induced by quasi-static shear or compression. The exponent \(\tau\) for these avalanches typically falls within the range of \(1.0\) to \(2.0\)~\cite{combe2000strain, sethna2001crackling, maloney2006amorphous, karmakar2010statistical, salerno2012avalanches, otsuki2014avalanche, denisov2016universality, shang2020elastic, oyama2021unified}.

In simulations, local minima can be  efficiently located using standard energy minimization algorithms such as gradient descent algorithm (GDA) or the FIRE algorithm~\cite{bitzek2006structural}. In contrast, searching for saddle points is considerably more challenging. The instantaneous normal mode approach, for instance, proposes estimating barrier heights from the spectrum of the Hessian matrix for instantaneously sampled configurations~\cite{keyes1997instantaneous, doye2002saddle, broderix2000energy, angelani2000saddles}. However, the general validity of this correspondence has been questioned~\cite{gezelter1997can}. An alternative method involves minimizing the auxiliary function \( |\nabla E|^2 \)~\cite{angelani2000saddles, broderix2000energy}, yet many of its minima do not correspond to true stationary points on the energy landscape~\cite{doye2002saddle}. Other specialized algorithms have been developed, including the numerical polynomial homotopy continuation method~\cite{mehta2011homotopy}, the deflation technique~\cite{farrell2015deflation}, the WKBJ-based approach~\cite{2020FromWKBJ}, and Systematic Excitation ExtRaction algorithm~\cite{pica2024local}. A common limitation of these conventional methods~\cite{angelani2000saddles, broderix2000energy, doye2002saddle, mehta2011homotopy, farrell2015deflation, 2020FromWKBJ} is their sensitive dependence on suitable initial guesses. Furthermore, they do not inherently reveal the topological relationships between different stationary points.

Here we employ the recently developed saddle dynamics algorithm (SDA)~\cite{yin2020construction} to study the complex energy landscape of a standard granular model. A key advantage of this method is its ability to efficiently construct a {\it pathway map} that organizes stationary points according to their Morse indices~\cite{1989Morse}. The SDA has been successfully applied to several energy-based systems with continuous variables, including the Ginzburg--Landau model for phase transitions~\cite{yin2021searching}, the Landau--de Gennes model for liquid crystals~\cite{yin2020construction}, the Gross--Pitaevskii model for Bose--Einstein condensation~\cite{yin2022constrained}, and the Lifshitz--Petrich model for quasicrystals~\cite{Yin2021transition}. The pathway map given by the SDA provides a convenient mathematical tool for understanding the complex organization of minima and saddles. Using this approach, we demonstrate directly that energy barriers are hierarchically and ultrametrically organized near the jamming transition.

{\bf Pathway map of stationary points.} 
To model jammed granular matter, we simulate packings of $N$ frictionless, monodisperse soft spheres with harmonic and Hertzian contact interactions, under periodic boundary conditions (see details in SM Sec.~S2). 
The density (volume fraction) of the generic jamming transition (jamming-point density) is $\varphi_{\rm J} \approx 0.64$~\cite{o2003jamming}. 
The potential energy landscape (PEL) of this athermal model is expected to be hierarchical and ultrametric near the jamming transition, according to the mean-field replica theory~\cite{Parisi_Urbani_Zamponi_2020}. The $\bm{x}$ in ${E}(\bm{x})$ is a $3N$-dimensional vector that stands for particle coordinates.

In the standard GDA, the search flow $\dot{\boldsymbol{x}} =-\nabla {E}(\bm{x})$ on the PEL generally brings the system state to the nearest minimum. The spirit of the SDA is to revert the search flow in the direction of Hessian eigenvectors with the $k$ smallest eigenvalues, in order to search for a {\it $k$-saddle}. The Morse index~\cite{1989Morse}, i.e., the number of negative eigenvalues of the Hessian $H = \nabla^2 E(\bm x)$, of a $k$-saddle is $k$. According to this definition, a local minimum is a 0-saddle. To enhance computational efficiency, we have implemented momentum-based acceleration algorithms in numerical discretization \cite{luo2025acc} (see SM Secs.~S3 and~S4).

 Our numerical strategy  is illustrated by the pathway map in Fig.~\ref{fig1}b. We take a simple example of a two-variable $\bm{x}= (x, y)$ energy function {$E(\bm{x})$}, illustrating its stationary points and the SDA search pathways.  If a $(k-1)$-saddle is found via the downward search from a $k$-saddle (vice versa in a upward search), they are connected by a pathway, with the arrow indicating the direction of a descending $k$. The pathway map of $E(\bm{x})$ can be also represented by a tree-like graph organizing stationary points in a descending order of $k$, with leaf nodes being the local minima (see Fig.~\ref{fig1}c).

We independently generate  4-saddles by SDA with the upward search strategy from random initial minima, and then construct a pathway map from each 4-saddle using the SDA with the downward search strategy (see SM Secs.~S5 and~S6).
The volume fraction $\varphi$ is fixed in the SDA and is kept above  $\varphi_{\rm J} \approx 0.64$. 
The 4-saddles at the root are independent at different $\varphi$, and the statistics are carried out for the stationary points at the given $\varphi$ (or pressure $p$). 
Due to the limitation in the  computational efficiency, the current approach cannot exhaustively enumerate all lower-rank saddles connected to the given 4-saddle
(the number of stationary points grows exponentially with $N$), 
but the statistical results are representative and do not sensitively depend on the number of sampled  stationary points.

\begin{figure}
    \centering
    \includegraphics[width=1\linewidth]{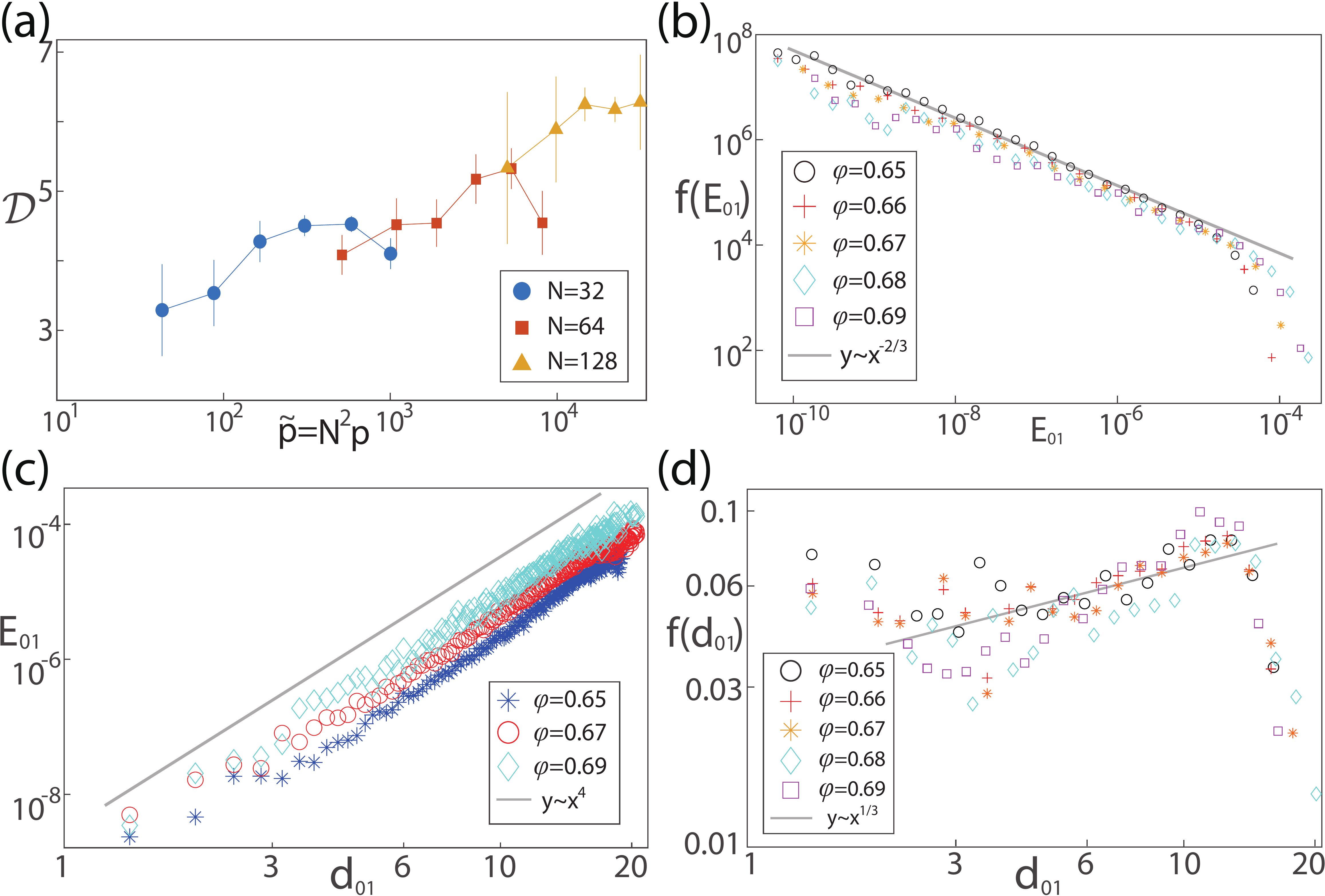}
    \caption{{\bf Statistical properties of distances and energy barriers measured from the native pathway map (without reconstruction).}
    (a) The deviation of $d$ from perfect ultrametricity, $\mathcal{D}(\tilde{p})$. Data are obtained for particles with Harmonic interactions.
    (b) The probability density $f(E_{01})$ follows a power law with an exponent of about −2/3.
    (c) The relation between $E_{01}$ and  $d_{01}$ reveals a power law with an exponent of about 4.
    (d) The probability density of $d_{01}$ compared to $f(d_{01}) \sim d_{01}^{1/3}$.
    Data in (b-d) are obtained for Hertzian systems with $N=64$ particles.
    }
    \label{fig2}
\end{figure}

\begin{figure*}
    \centering
    \includegraphics[width=0.9\linewidth]{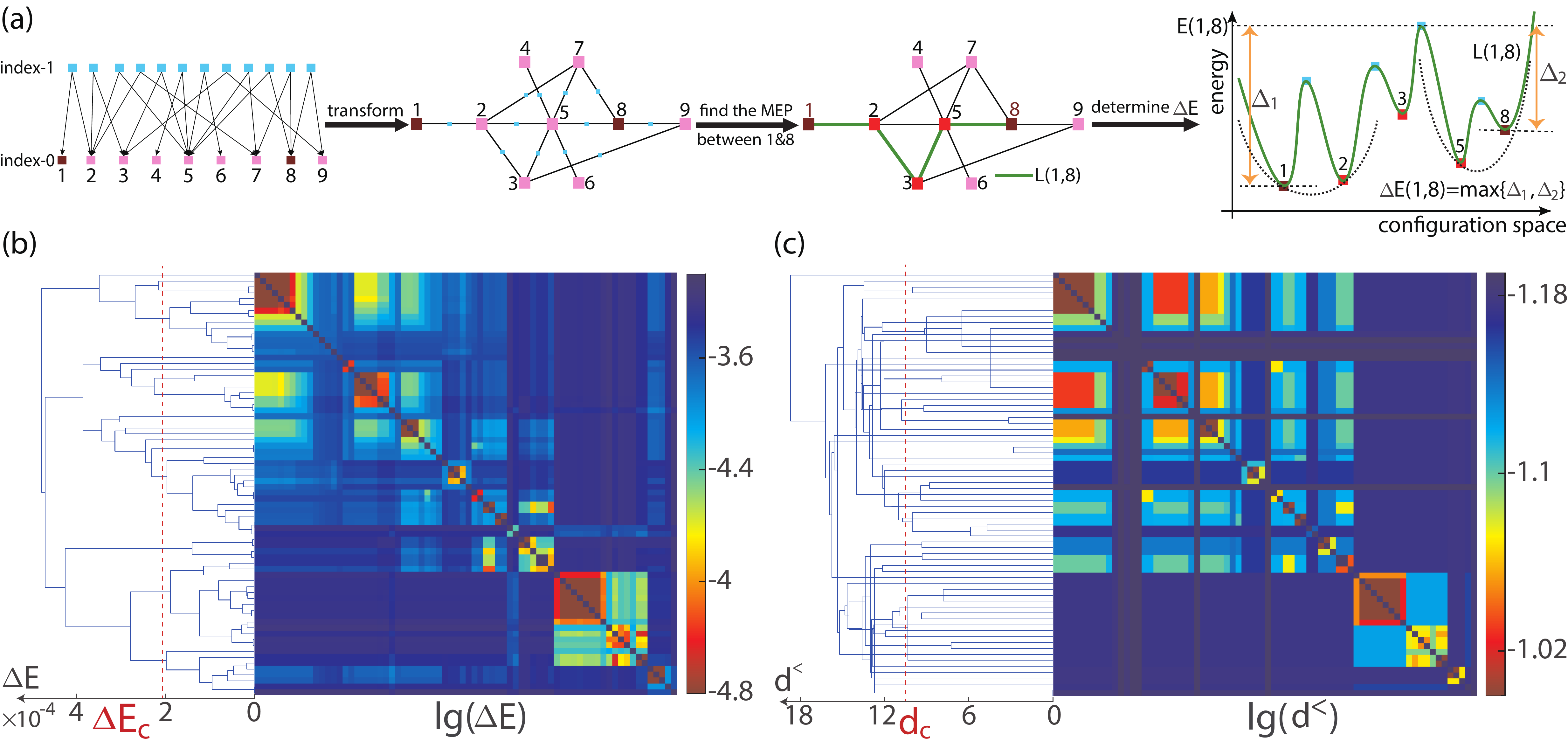}
\caption{{\bf Reconstruction of the pathway map with minimum energy pathways.}
 (a) A schematic example of finding the min-max pathway $L_{\rm max}^{\rm min}(1,8)$ between two minima (1,8) from the native pathway map (from left to right).
The matrix and tree representations of (b) $\Delta E$ and (c) $d^{<}$ on a set of local minima, organized using the clustering algorithm. The color scales with the corresponding value.
}
\label{fig3}
\end{figure*}

{\bf Hierarchy and ultrametricity of inter-minimum distances.}
We first confirm that the inter-minimum (0-0 saddle pairs) distance obtained by the SDA is hierarchical and ultrametric, consistent with the mean-field theory~\cite{Parisi_Urbani_Zamponi_2020}
and previous numerical study~\cite{dennis2020jamming}.
The inter-minimum  distance can be defined in various ways, e.g., the square distance~\cite{charbonneau2015numerical, berthier2016growing, artiaco2020exploratory} or overlap~\cite{ninarello2017models,  artiaco2020exploratory} between the two configurations, or the inherent structure minimal displacement achieved by  pattern-matching~\cite{yu2024universal}.  
To emphasize that the complexity of jamming landscape is originated from mechanical isostaticity and marginality, we follow the {\it distance  metric} defined in ~\cite{dennis2020jamming} for a pair $(a,b)$ of minima,
\begin{equation} \label{d metric}
    d({a},{b}) = \frac{1}{\sigma} \sqrt{\sum_{i j}\left(\vec{C}_{{a}}^{i j}-\vec{C}_{{b}}^{i j}\right)^2},
\end{equation}
where $\vec{C}_{{a}}^{i j}$ denotes the stable contact vector between particles $i$ and $j$, and $\sigma$ the diameter of a particle. The contact vector $\vec{C}_{{a}}^{i j}$ is taken as $ \bm{x}_i - \bm{x}_j $ if two particles are in contact, otherwise $\bm{0}$.  
Following the strategy in~\cite{dennis2020jamming}, we confirm that $d$ is ultrametric in the thermodynamic limit. Ultrametricity means that the distances  between any three configurations, $a,b$ and $c$, satisfy the strong triangle inequality,
\begin{equation}
d(a,c) \leq \mathop{\max} [d(a,b), d(b,c)].
\label{eq:ultrametric_inequality}
\end{equation}

The basic idea is as follows. First, using a minimum spanning tree method~\cite{MST01, MST02, MST03, MST04, MST05}, one constructs the so-called subdominant ultrametric $d^<$, which is the perfect ultrametric lower than and closest to the original metric $d$ (see SM Sec.~S7). In geneal, $d({a},{b})$ and $d^<({a},{b})$ are different because $d$ is not perfectly ultrametric. However, if one can show that the difference between $d$ and $d^<$ vanishes as the scaled pressure $\tilde{p} = N^2 p \to 0$,  then ultrametricity is established in the thermodynamical limit $N \to \infty$ near the jamming transition. Here $\tilde{p}$ is interpreted as the distance to jamming, taking into account the established finite-size scaling $p \sim N^{-2}$ (for harmonic interactions)~\cite{goodrich2012finite}. 
The above difference is quantified by 
$
    \mathcal{D} \equiv \sqrt{\left\langle\left [ d(a, b)-d^{<}(a, b)\right ]^2\right\rangle},
$
where the angle bracket denotes an average over all pairs of minima $(a,b)$ in the constructed pathway map. 
The dimensionless difference is $ \mathcal{D}/ d$, where the typical distance  $d \sim \sqrt{Z N}$ with the coordination number $Z \approx 6$ near jamming~\cite{alexander1998amorphous}. Thus one should consider a proper scaled difference  $\tilde{\mathcal{D}} \equiv  \mathcal{D}/\sqrt{N} \sim \mathcal{D}/ d$. Fig.~\ref{fig2}a shows that the data of $\mathcal{D}$ for different $N$ and pressure $p$ collapse, when plotted as a function of $\tilde{p}$. The master curve of $D(\tilde{p})$ implies that $D(\tilde{p} \to 0) \approx 3$, or
$\tilde{\mathcal{D}}(\tilde{p}\to 0) \to 0$, and therefore $d$ is ultrametric in the thermodynamic limit. 

Using this strategy, Ref.~\cite{dennis2020jamming} has shown the ultrametricity of $d$ near jamming. We emphasize that the local minima are sampled in different ways.
In this work, the minima are all inter-connected through higher-index saddles, sampled by the SDA. In contrast, in Ref.~\cite{dennis2020jamming}, the minima are sampled from repeated random perturbations of the system, followed by re-minimizing of the energy that brings the system to the nearby stable configuration. The consistent results obtained by the two numerical methods establish robustly ultrametricity, and validate the current SDA approach. Importantly, saddles cannot be sampled using the method in Ref.~\cite{dennis2020jamming}.

Fig.~\ref{fig3}c visualizes the hierarchical structure of the distance metric. The minima are hierarchically clustered using MATLAB's {\it dendrogram} function based on the subdominant ultrametric $d^<$, and the matrix of $d^{<}$ after clustering is visualized (see also SM Sec.~S8). 
The minima with the smallest $d^{<}$ are firstly grouped together, and then grouped with other minima consecutively  with the ascending $d^{<}$. This procedure is represented by a tree in Fig.~\ref{fig3}c. Our numerical result is consistent with the full-RSB picture: the minima with the smallest $d^{<}$ form  ``basins'' that define glass states, and then form meta-basins for states with a larger $d^{<}$, and meta-meta-basins with an even larger $d^{<}$, etc.

\begin{figure}
    \centering
    \includegraphics[width=1\linewidth]{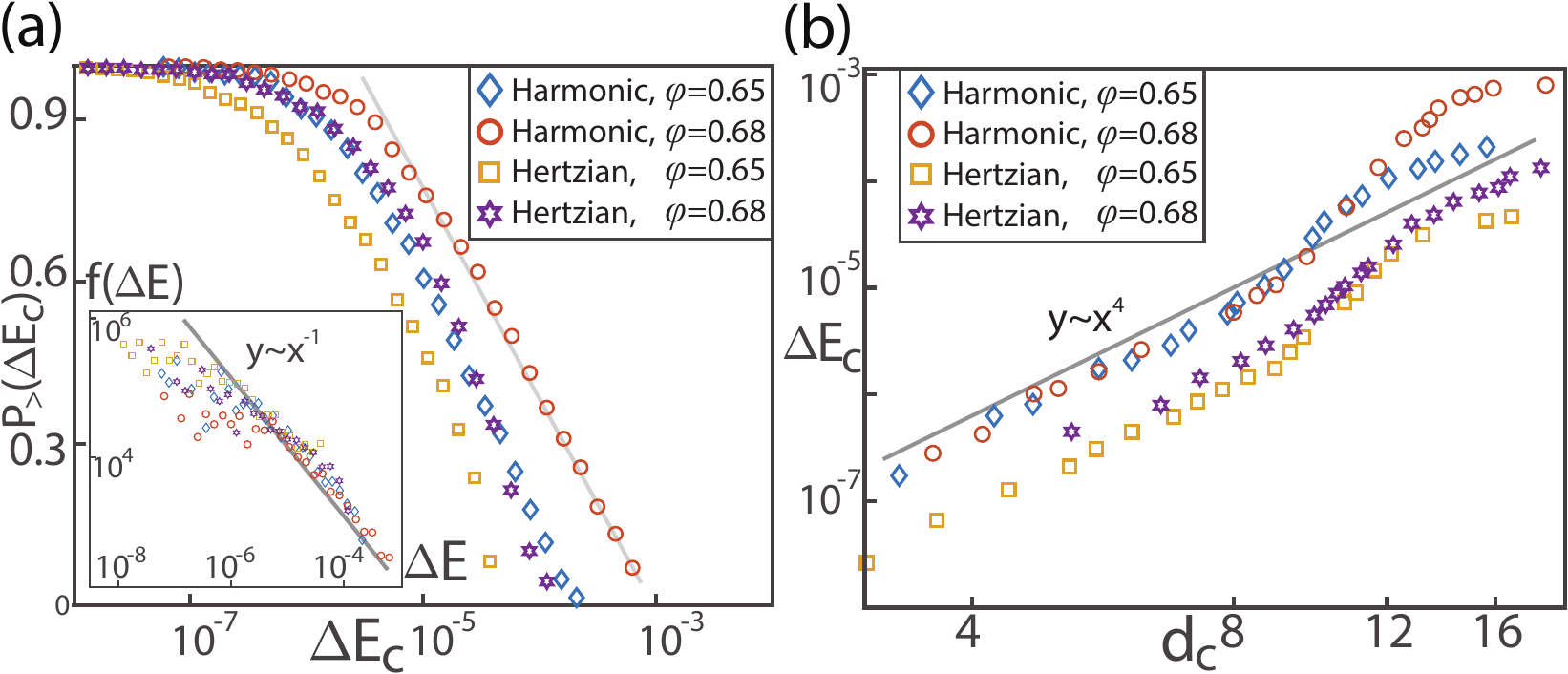}
\caption{
  {\bf Statistical properties of $\Delta E$.}
    (a) Data of  $P_>(\Delta E_{\rm c})$ and (inset) $f(\Delta E)$,  for systems with $N=64$ particles.
    (b) Correspondence between $\Delta E_{\rm c}$ and $d_{\rm c}$. 
}
    \label{fig4}
\end{figure}

{\bf Statistics of energy barriers and distances between minimum-saddle pairs.}
Next we explore the near-jamming statistics of energy barriers.
We first consider the energy difference $E_{01}$ between any 
 directly connected minimum-(1-saddle) pairs  in the pathway map (0-1 pairs for short). Fig.~\ref{fig2}b shows $f(E_{01}) \sim E_{01}^{-\tau}$ under different $\varphi$, with 
 {$\tau = 0.648 \approx 2/3$} from fitting (see SM Sec.~S9 for additional results on Harmonic systems of  different $N$).  
 The measured exponent $\tau \approx 2/3$ is clearly distinct from the mean-field exponent $\tau_{\rm MF} = 3/2$ and most numerical values $1<\tau<2$ of avalanche distributions. Physically, the considered $E_{01}$ is associated to a single 0-1 barrier crossing event, while an avalanche is typically formed by a sequence of barrier crossing events due to collective motions.
  Crossing over higher-index saddles is possible but rare, because the saddle energy monotonically increases with $k$ \cite{pechukas1981transition, truhlar1984variational, weinan2002string}. 
 Statistical properties for events crossing higher-index saddles ($k\geq 2$) are very interesting, but will be left for future studies.

To investigate the relationship between $d_{01}$ and $E_{01}$, we collect the $(d_{01}, E_{01})$ data of all direct 0-1 pairs. 
Here $d_{01}$ is the distance between the corresponding 0-1 pair, which shall not be confused with the 0-0 distance metric $d$ analyzed above.
Fig.~\ref{fig2}c shows a power-law correlation, $E_{01} \sim d_{01}^ \lambda$, where $\lambda = 4.03 \approx 4$ from fitting.  Since the  two distributions are related by a transformation $f(E_{01})d(E_{01})=f(d_{01})d(d_{01})$, $f(d_{01})$ should follow a power law distribution with an exponent of about $1/3$, confirmed by  the data (see Fig.~\ref{fig2}d). 
Explanations for the observed scalings, $E_{01} \sim d_{01}^4$ and $f(E_{01}) \sim E_{01}^{-2/3}$, are provided in SM Secs.~S10 and~S11.

{\bf Hierarchy and ultrametricity of inter-minimum energy barriers.}  
While $E_{01}$ analyzed above quantifies the energy difference between a directly connected minimum-saddle pair, we are also interested in  the energy cost  $\Delta E$ of the transitional path between two arbitrarily separated minima. 
For this purpose, a reconstruction of the pathway map is needed, to select the {\it minimum energy pathway} (MEP) out of all possible pathways between two minima. This
is based on the consideration that the probability of a certain activated pathway decreases exponentially with an increasing energy cost, according to the Arrhenius law. Thus only the pathway with the lowest energy cost is essential~\cite{fontanari2002fractal,FLAMM_FONTANA_2000RNA,hordijk2003shapes, li2024simplest}.

In general, two minima can be connected through multiple 0-1-0-$\cdots$-1-0 pathways. The collection of such pathways is denoted as $\mathscr{L} = \{ L \}$.
Each pathway $L$ contains a series of stationary points $c_1^L, c_2^L, \ldots $ between minima $a$ and $b$, with the maximum energy $E_{\rm max}^L(a,b) = \mathop{\max}\limits_i { E(c_i^L)}$. We define MEP as the pathway with the minimum $E_{\rm max}^L(a,b)$, characterized by $L_{\rm max}^{\rm min}(a,b) \equiv \mathop{\arg\min}\limits_{L\in \mathscr{L}} E_{\rm max}^L(a,b)$, with the min-max energy $E_{\rm max}^{\rm min}(a,b) \equiv \mathop{\min}\limits_{L\in \mathscr{L}} E_{\rm max}^L(a,b)$ (or simply denoted as $L(a,b)$ and $E(a,b)$).  
The energy barrier $\Delta E$ is then defined by $\Delta E(a,b) \equiv \mathop{\max}[E_{\rm max}^{\rm min}(a,b)-E(a),E_{\rm max}^{\rm min}(a,b)-E(b)]$. 
Fig.~\ref{fig3}a shows the example of calculating $\Delta E(1,8)$ starting from the native pathway map. 
Our strategy is essentially similar to the ``min-max'' formulation used for spin glasses ~\cite{fontanari2002fractal,FLAMM_FONTANA_2000RNA,hordijk2003shapes, li2024simplest}.

By construction, $E_{\rm max}^{\rm min}(a,b)$ satisfies Eq.~(\ref{eq:ultrametric_inequality}), and therefore is ultrametric.
An ultrametric measure is unnecessarily hierarchical; e.g., a one-level tree is ultrametric (a tree is always ultrametric) but not hierarchical.
However, our analysis reveals that $\Delta E$ is indeed hierarchical near jamming. This can be seen from the clustering results in Fig.~\ref{fig3}b, which resembles the hierarchical structure of the corresponding $d^<$ (Fig.~\ref{fig3}c).

The hierarchy of $\Delta E$ is revealed quantitatively by a scale-free (power-law) distribution $f(\Delta E)$. The data of the complementary cumulative distribution function (CCDF), defined as $P_>(\Delta E_{\rm c}) = 1-\int_0^{\Delta E_{\rm c}} f(\Delta E) d (\Delta E)$, shows a logarithmic tail, $P_>(\Delta E_{\rm c}) \sim \ln (\Delta E_{\rm c})$, which means $f(\Delta E) \sim \Delta E^{-1}$ (see Fig.~\ref{fig4}a). Interestingly, The exponent $\tau =1$ coincides with the full-RSB theoretical result for over-jammed soft spheres, $\tau_{\rm UNSET} = 1$~\cite{franz2017mean}.
To further establish the relation between energy barriers and distances, we calculate the correspondence between $\Delta E_{\rm c}$ and $d_{\rm c}$ satisfying $P_>(\Delta E_{\rm c}) = P_>(d_{\rm c})$, where $P_>(d_{\rm c})$ is the CCDF of $d^{<}$ . 
Fig.~\ref{fig4}b shows that $\Delta E_{\rm c} \sim (d_{\rm c})^{\lambda'}$, with {$\lambda' = 4.23 \approx 4$} from fitting.
The relationship $\Delta E_{\rm c} \sim (d_{\rm c})^{4}$ is akin to $E_{01} \sim d_{01}^{4}$ obtained above.

{\bf Conclusion.} The current SDA-based approach opens the door for investigating a wide range of 
complex systems, such as polymers~\cite{xu2022thermodynamic}, proteins~\cite{hu2016dynamics}, 
and deep neural networks~\cite{yoshino2020complex, huang2025liquid}, all of which exhibit 
signs of  a multi-scale, hierarchical landscape. \\

{\bf Acknowledgments.}
Y. J. acknowledges funding from National Key R\&D Program of China (Grant No. 2025YFF0512000),
Wenzhou Institute (No. WIUCASICTP2022) and the National Natural Science Foundation of China (No. 12447101). 
L.Z. was supported by the National Natural Science Foundation of China (No. 12225102, T2321001, and 12288101) and the National Key R\&D Program of China 2024YFA0919500.
D.P. acknowledges funding from the National Natural Science Foundation of China (No. 12404290).
We acknowledge the use of the High Performance Cluster at Institute of Theoretical Physics, Chinese Academy of Sciences. 
\clearpage

\bibliographystyle{apsrev4-1}

\bibliography{ref}

\end{document}


\maketitle

\setcounter{figure}{0}
\setcounter{equation}{0}
\setcounter{table}{0}
\setcounter{section}{0}
\renewcommand\thefigure{S\arabic{figure}}
\renewcommand\theequation{S\arabic{equation}}
\renewcommand\thesection{S\arabic{section}}
\renewcommand\thetable{S\arabic{table}}

\tableofcontents
\section{Example of the energy landscape}
\begin{figure}[htb]
    \centering
    \includegraphics[scale=0.8]{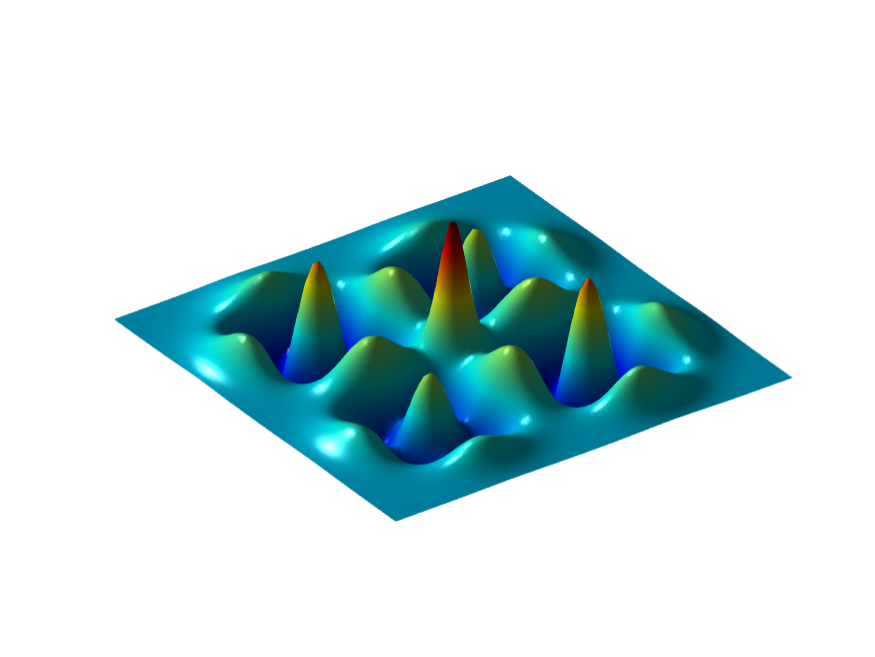}
    \caption{{\bf Three-dimensional view of the energy landscape $E(x,y)$}
    }
    \label{smfig1}
\end{figure}

The example $E(x,y)$ in Fig.~1b of the main text is a smooth two‑dimensional scalar function formed by summing Gaussian‑envelope polynomial components placed in a symmetric arrangement around the origin, together with a central isotropic Gaussian term:
 \begin{equation}
\begin{aligned}
f_1(x,y) &= 1.3 \left( \big((x-a)^2 - 1\big)^2 + \big((y-a)^2 - 1\big)^2 - 1.5 \right) \exp\left( -(x-a)^2 - (y-a)^2 \right),\\
f_2(x,y) &= 0.8 \left( \big((x+a)^2 - 1\big)^2 + \big((y-a)^2 - 1\big)^2 - 1.5 \right) \exp\left( -(x+a)^2 - (y-a)^2 \right),\\
f_3(x,y) &= 0.7 \left( \big((x-a)^2 - 1\big)^2 + \big((y+a)^2 - 1\big)^2 - 1.5 \right) \exp\left( -(x-a)^2 - (y+a)^2 \right),\\
f_4(x,y) &= 1.2 \left( \big((x+a)^2 - 1\big)^2 + \big((y+a)^2 - 1\big)^2 - 1.5 \right) \exp\left( -(x+a)^2 - (y+a)^2 \right),     \\
f_0(x,y) &= \exp \left( -6 (x^2 + y^2) \right),   \\ 
E(x,y) & = f_1(x,y) + f_2(x,y) + f_3(x,y) + f_4(x,y) + f_0(x,y),
\end{aligned}
 \end{equation}

where the parameter $a=2$. The energy landscape is depicted in Fig.~\ref{smfig1}.

\section{Granular matter model}
The systems consist of $N$ identical, frictionless grains with diameter, $\sigma$.
The box is cubic and the periodic boundary is used. 
Grains interact with each other if the center-to-center distance is less than $\sigma$ and the pair potential is $V(r_{\rm ij}) = \frac{1}{\alpha}\varepsilon(1 - \frac{r_{\rm ij}}{\sigma})^{\alpha}$. The exponent $\alpha$ is $5/2$ and $2$ for Hertzian and harmonic interaction, respectively.
The diameter $\sigma$ and interaction energy scale $\varepsilon$ are used as the length and energy units.

\section{Saddle dynamics algorithm and downward/upward search}
The algorithm searching for saddles of the energy landscape ${E}(\bm{x})$ consists of saddle dynamics algorithm (SDA, also known as the high-index saddle dynamics algorithm, HiSD) and a downward/upward search algorithm.
For a given integer $k \geq 0$, the $k$-index saddle dynamics ($k$-SD) flow designed to search for $k$-saddles is a modification of the gradient descent (GD) flow 
\begin{equation}
\label{eq:GD}
\dot{\bm{x}} =-\nabla {E}(\bm{x}),
\end{equation}
with additional contributions from the $k$ vectors depicting the $k$-dimensional unstable eigen-subspace of the Hessian $H=\nabla \nabla E$:
\begin{equation} \label{eqn:SDA1}
\left\{
\begin{aligned}
\dot{\bm{x}}  &=-\left(\mathbf{I}-\sum\limits_{i=1}^{k} 2\bm{v}_{i} \bm{v}_{i}^{\top}\right) \nabla E,\\
\dot{\bm{v}_i}&=-\left(\mathbf{I}-\bm{v}_i\bm{v}_i^\top- \sum\limits_{j=1}^{i-1} 2\bm{v}_j\bm{v}_j^\top\right) H \bm v_i ,\quad i=1, \cdots k, 
\end{aligned}
\right.
\end{equation}
where $\bm{x}$ and $\{ {\bm{v}}_{i} \}_{i=1}^k$ are treated as functions of time $t$. The dynamics of $\{ {\bm{v}}_{i} \}_{i=1}^k$ identifies, at any given time $t$, the eigenvectors corresponding to $k$ negative eigenvalues of Hessian $H(\bm{x})$, maintaining the orthonormal constraints $\bm{v}_i \cdot \bm{v}_j=\delta_{ij}$. 

To complete the dynamics, the initial conditions, $\bm{x}_0 \equiv \bm{x}(t=0)$ and $\{\bm{v}_i^0\}_{i=1}^k \equiv \{\bm{v}_i(t=0)\}_{i=1}^k$ are required. The orthonormal vectors $\{\bm{v}_i^0\}_{i=1}^k$ are usually taken as the smallest $k$ eigenvalues of Hessian $H(\bm{x}_0)$. Starting from a proper initial guess, a $k$-saddle $\bm{x}^*$ along with its $k$ unstable eigenvectors $\{ {\bm{v}}_{i}^* \}_{i=1}^k$ will be attained. 
Given an initial guess, the $k$-saddles connected to it can be locally identified for any integer $k$, whenever they exist. To construct a large-scale network using SDA, nonlocal methods should be employed.
The downward (upward) search algorithm is initialized at a saddle and, with an appropriate perturbation, seeks connected lower (higher) index saddles.

The downward search strategy is to apply $k$-SDA starting from a parent $m$-saddle $\bm{x}^*$ to search $k$-saddle ($k<m$). 
The starting point of SDA, $\bm{x}_0=\bm{x}^* \pm \delta \bm{x}$, is set to push the system away from $\bm{x}^*$. The pushing direction $\delta \bm{x}$ is along a linear combination of $(m-k)$ vectors chosen from $\{ {\bm{v}}^*_{i} \}_{i=1}^m$ (the eigenvectors of $H$ at $\bm{x}^*$) whose negative eigenvalues have the smallest magnitudes. 
A preliminary attempt to choose $\hat{{\bm{v}}}$ is to try: $\delta \bm{x}=\epsilon \bm{v}_{i}, i=1,\cdots m$, and set the initial state of $k$-SDA as $\left(\bm{x}^* \pm \epsilon \bm{v}_{i}^*; \bm{v}_{1}^*, \cdots, \bm{v}_{k}^*\right)$, if $i>k$; $\left(\bm{x}^* \pm \epsilon \bm{v}_{i}^*; \bm{v}_{1}^*, \cdots, \bm{v}_{i-1}^*,\bm{v}_{i+1}^*,\cdots, \bm{v}_{k+1}^*\right)$, if $i \leq k$ for small $\epsilon$. Intuitively, this strategy makes the system to stay unstable in the chosen $k$ directions but to roll off the original $m$-saddle in other $(m-k)$ directions containing the perturbation direction $\delta \bm{x}$. 
Once a new index-$k$ saddle point is found (i.e. the $k$-SDA flow Eq.~\ref{eqn:SDA1} converges), an arrow from the $m$-saddle $\bm{x}^*$ to this stationary state is drawn in the pathway map to illustrate this downward relationship.

We can also use upward search when the parent state is unknown or we need the information of higher-index saddles based on the existing minima or saddles. The search for $m$-saddles $(m>k)$ starting from a $k$-saddle $\bm{x}^*$ can be implemented as follows. Choose a stable direction $\hat{{\bm{v}}}$ and $m$ orthonormal initial directions including $\hat{{\bm{v}}}$ at $\bm{x}^*$. A typical choice for $m$-SDA upward search initiation is $\left(\bm{x}^* \pm \epsilon \bm{v}_{m}^*; \bm{v}_{1}^*, \cdots, \bm{v}_{m}^*\right)$. 

It has been proved that $(\bm{x}^*, \bm{v}_{1}^*, \cdots, \bm{v}_{k}^*)$ is a linearly stable stationary point of the $k$-SD if and only if $\bm{x}^*$ is a $k$-saddle of the system ~\cite{yin2021searching}. Thus the downward and upward search algorithms can reach saddles of the specified order $k$ if $k$-SDA converges.

To construct the pathway map for our granular model, we first randomly select a 4-saddle, which is generated by performing an upward search starting from a randomly chosen local minimum. Subsequently, a downward search is conducted to identify 3-saddles connected to the parent 4-saddle. Each identified 3-saddle is then treated as a new initial saddle, and the downward search process is repeated iteratively until all leaf nodes are local minima.

\section{Numerical discretization of the saddle dynamics algorithm and acceleration}
In numerical implementation, the dynamics of $k$-SDA (Eq.~\ref{eqn:SDA1}) is discretized in time using explicit Euler schemes on $\bm{x}^{(n)}$ ($n$ denotes the discrete time step) and semi-implicit scheme on $\{ {\bm{v}}_{i}^{(n)} \}_{i=1}^{k}$:
\begin{equation} \label{eqn:SDA2}
\left\{
\begin{aligned}
\bm{x}^{(n+1)} &= \bm{x}^{(n)} - \alpha \left(\mathbf{I} - 2\sum_{i=1}^{k} \bm{v}_i^{(n)} {\bm{v}_i^{(n)}}^\top\right)\nabla E\left( \bm{x}^{(n)} \right ),\\
\bm{v}_i^{(n+1)} & = \bm{v}_i^{(n)}  -\beta \left(\mathbf{I}-\bm{v}_i^{(n)}{\bm{v}_i^{(n)}}^\top- \sum\limits_{j=1}^{i-1} 2\bm{v}_j^{(n)}{\bm{v}_j^{(n)}}^\top\right) H\left(\bm{x}^{(n+1)}\right) \bm v_i^{(n)} ,\quad i=1, \cdots k. 
\end{aligned}
\right.
\end{equation}
The update rates $\alpha$ and $\beta$ can be configured as either Barzilai-Borwein (BB) step sizes~\cite{1988Two} or fixed step sizes.
The explicit form of $F=\nabla E$ and $H$ can be obtained by differentiating energy $E(x)$ with respect to $x$. Practically we compute the matrix-vector multiplication in each iteration step as $H\bm{v}_i=(\bm{F}\left(\bm{x}+\bm{l} \bm{v}_{i}\right)-\bm{F}\left(\bm{x}-\bm{l} \bm{v}_{i}\right))/2 l$ by a dimer method with a small constant $l>0$. This approximation does not affect the SDA's error or convergence, but makes it more efficient. SDA does not necessarily require a closed form of the Hessian, which makes this method more applicable compared to other complicated models.

As the initiation of SDA, the initial eigenvectors $\{\bm{v}_i^*\}_{i=1}^{k}$ of $H$ (explicit form) at the starting point $\bm{x}^*$ are efficiently and accurately solved by the Locally Optimal Block Preconditioned Conjugate Gradient algorithm (LOBPCG) ~\cite{KnyazevLOBPCG01}. 

To enhance computational efficiency, the accelerated high-index saddle dynamics (A-HiSD) ~\cite{luo2025acc} is adopted in the iterative steps within the convergence region of the $k$-SDA flow, and it is numerically realized by adding a term $\Gamma^{(n)}=\gamma \left(\bm{x}^{(n)} - \bm{x}^{(n-1)}\right)$ in the update process from $\bm{x}^{(n)}$ to $\bm{x}^{(n+1)}$:
\begin{equation} \label{eqn:ASDA}
\left\{
\begin{aligned}
\bm{x}^{(n+1)} &= \bm{x}^{(n)} - \alpha \left(\mathbf{I} - 2\sum_{i=1}^{k} \bm{v}_i^{(n)} {\bm{v}_i^{(n)}}^\top\right)\nabla E\left(\bm{x}^{(n)}\right) + \gamma \left(\bm{x}^{(n)} - \bm{x}^{(n-1)}\right),\\
\bm{v}_i^{(n+1)} & = \bm{v}_i^{(n)}  -\beta \left(\mathbf{I}-\bm{v}_i^{(n)}{\bm{v}_i^{(n)}}^\top- \sum\limits_{j=1}^{i-1} 2\bm{v}_j^{(n)}{\bm{v}_j^{(n)}}^\top\right) H\left(\bm{x}^{(n+1)}\right) \bm v_i^{(n)} ,\quad i=1, \cdots k. 
\end{aligned}
\right.
\end{equation}
The added term $\Gamma$ is a momentum-based term inspired by the simple implementation and the efficient acceleration phenomenon of the heavy-ball method ~\cite{POLYAK19641} in GDA.
In practice, during each iteration process of $k$-SDA, if the norm of the gradient of $E$ (i.e., $\| \nabla E \|$) decreases to a value below $\epsilon_1$, the update rate of $\bm{x}$, $\alpha$ and $\gamma$, are set to be:
\begin{equation} \label{eqn:alphagamma}
\alpha=\left( \frac{2}{\sqrt{\mathop{\max}_i{|\lambda_i|}}+\sqrt{\mathop{\min}_i{|\lambda_i|}}} \right)^2 , \quad \gamma=\left( 1-\frac{2}{\sqrt{\kappa}+1} \right)^2 , 
\end{equation}
where $\{ \lambda_i \}$ is the set of eigenvalues of $H$, and $\kappa = \mathop{\max}_i{|\lambda_i|}/\mathop{\min}_i{|\lambda_i|} $ is the 2-norm condition number of $H$. When $\| \nabla E \|>\epsilon_1$, $\alpha$ is the BB step size and $\gamma$ is set to $0$, and thus the numerical scheme Eq.~\ref{eqn:ASDA} reduces to Eq.~\ref{eqn:SDA2}. The algorithm is outlined in Algorithm [1], and the parameters are given in Table~\ref{tab:parameters}.

\begin{algorithm}[H]
    \caption{$k$-SDA with acceleration}
    \begin{algorithmic}[1]
        \STATE \textbf{Input:} $k \in \mathbb{N}, \bm{x}^{(0)} \in \mathbb{R}^d$, orthonormal vectors $\{\bm{v}_i^{(0)}\}_{i=1}^k \subset \mathbb{R}^d$, $\epsilon_1, \epsilon_2 > 0$, $N_{\rm max} \in \mathbb{N}$.
        \STATE Set $\bm{x}^{(-1)} = \bm{x}^{(0)}$, $n = 0$.
        \WHILE{$n < N_{\rm max}$}
            \IF{$\| \nabla E(\bm{x}^{(n)})\| < \epsilon_2$}
                \STATE \textbf{return} $\bm{x}^{(n)}$, 
                $\{\bm{v}_i^{(n)}\}_{i=1}^k$;
            \ENDIF
            \IF{$\| \nabla E(\bm{x}^{(n)})\| > \epsilon_1$}
                \STATE $\bm{x}^{(n+1)} = \bm{x}^{(n)} - \alpha  \left( I - 2 \sum_{i=1}^k \bm{v}_i^{(n)} \bm{v}_i^{(n)\top} \right) \nabla E(\bm{x}^{(n)})$;   ($\alpha$ takes BB step size or fixed step size)
            \ELSE
                \STATE $\Gamma^{(n)}=\gamma (\bm{x}^{(n)} - \bm{x}^{(n-1)})$;
                \STATE $\bm{x}^{(n+1)} = \bm{x}^{(n)} - \alpha  \left( I - 2 \sum_{i=1}^k \bm{v}_i^{(n)} \bm{v}_i^{(n)\top} \right) \nabla E(\bm{x}^{(n)}) + \Gamma^{(n)}$;
                ($\alpha$ and $\gamma$ are given by Eq. \ref{eqn:alphagamma})
            \ENDIF
            \STATE $\{\bm{v}_i^{(n+1)}\}_{i=1}^k = \text{EigenSol}\left( \{\bm{v}_i^{(n)}\}_{i=1}^k, \nabla^2 E(\bm{x}^{(n+1)}) \right)$; (the second equation in Eq. \ref{eqn:SDA2} or Eq. \ref{eqn:ASDA})
            \STATE $n = n + 1$;
        \ENDWHILE
        \STATE \textbf{return} \text{``Not Converge!''}
    \end{algorithmic}
\end{algorithm}

\begin{table}[htbp]
  \centering
  \caption{Parameter Values in numerical implementation of SDA}
  \begin{tabular}{ccl}
    \toprule
    Parameter & Value & Definition \\
    \midrule
    $N_{\rm max}$ &   $10^{5}$  &   Upper bound of iteration steps  \\
    $l$ & $10^{-6}$  & Dimer parameter in approximating $H\bm{v}_i$ \\
    $\epsilon_1$ & $10^{-11} \sim 10^{-10}$ & Threshold for introducing acceleration \\
    $\epsilon_2$ & $1 \times 10^{-13} \sim 3 \times 10^{-13}$ & Stopping criterion for iterations \\
    $\epsilon$ & $10^{-8} \sim 10^{-6}$ & Perturbation in the downward/upward search\\
    \bottomrule
  \end{tabular}
  \label{tab:parameters}
\end{table}

\section{Protocol to generate initial 4-saddles}
The initial 4-saddles are generated with the following procedure.
Firstly, we generate random configurations at a given packing fraction $\varphi$.
Then, the system is quenched into a nearby energy minimum state using the FIRE method~\cite{bitzek2006structural}.
Lastly, from that configuration, we employ the saddle dynamics algorithm with the upward search strategy to generate a 4-saddle.

\section{Removing degenerate minima}
Inside the simulation program, we set the box length as one to achieve best numerical stability, and other quantities are scaled according to this setup.
To avoid Hessian degeneracy induced by global drifts under periodic boundary conditions, one particle is fixed at the origin $(0,0,0)$; thus the configurational vector is $\bm{x} \in [-\frac{1}{2},\frac{1}{2}]^{3N-3}$.
To eliminate the non-determinacy in calculating the configurational distance between minima, minima that have (non-trivial) zero-energy  vibrational modes are 
considered have degeneracy and excluded from the subsequent analysis. 
We employ two different methods to identify degenerate minima： (i) the Hessian has non-trivial zero eigen-modes (the absolute value of the eigen-value is smaller than $10^{-12}$), (ii) minima that contain rattlers (rattlers are recursively defined as particles that contact fewer than $4$ non-rattler particles in three dimensions). We find that these two methods give almost the same result.

\section{Minimum spanning tree in calculating $d^<$ and minimum energy paths}

\subsection{Minimum spanning tree}
The \textbf{Minimum Spanning Tree (MST)} is a fundamental concept in the graph theory ~\cite{MST01, MST02, MST03, MST04, MST05}, specifically applied to \textbf{weighted connected undirected graphs}. Formally, given a weighted connected undirected graph $G = (V, E) $ — where $V$ denotes the set of vertices and $E$ denotes the set of edges with associated weights — an MST is a spanning tree of $G$ that satisfies two core properties:
i) It contains all vertices of $G$ and has exactly $n-1$ edges (where $n$ is the number of vertices in $V$ ), forming a connected acyclic subgraph;
ii) The sum of the weights of its edges is the smallest among all possible spanning trees of $G$.

We use the MST method to solve both the subdominant ultrametric $d^<$ and the MEP.

\subsection{Minimum spanning tree method used in calculating $d^<$}
For a given metric $d$, the corresponding subdominant ultrametric $d^<$ can be defined and derived through the following stepwise procedure ~\cite{dennis2020jamming}:
\begin{itemize}
    \item  [1)]
    First, the metric $d$ is represented as a symmetric matrix encoding pairwise distances between minima. This matrix can be equivalently mapped to an edge-weighted undirected graph, where vertices correspond to the minima and edge weights are defined as the pairwise distances $d$ between minima.
    \item  [2)]
    Next, we compute the MST of this graph by Kruskal's algorithm ~\cite{MST04} using MATLAB's {\it minspantree} function.
    Notably, a key property of this tree structure is that there exists exactly one simple path between any pair of vertices.
    \item  [3)]
    Finally, the subdominant ultrametric $d^<$ is constructed as a symmetric matrix. For any pair of vertices $i$ and $j$, the entry $d^{<}_{ij}$ in this matrix is determined by the maximum edge weight along the unique path connecting $i$ and $j$ in the aforementioned MST.
\end{itemize}

\subsection{Minimum spanning tree method used in solving the minimum energy path problem}
The MEP connecting two local minima $a,b \in S_0$ , denoted by $L_{\rm max}^{\rm min}(a,b)$ or simply $L(a,b)$, can be defined in a ``min-max'' formulation:
\begin{equation} \label{oriMEP1}
    L(a,b)=\mathop{\arg\min} \limits_{L\in \mathscr{L}} \mathop{\max} \limits_{z \in L} E(z),
\end{equation}
where $\mathscr{L} = \{ L \}$ denotes the set of 0-1-0-$\cdots$-1-0 pathways (multi-step transition processes) connecting $a,b \in S_0$.
The energy of the lowest 1-saddle $s_1(a,b) \in S_1$ that separates minima $a$ and $b$ is defined as:
\begin{equation} \label{oriMEP2}
    E(a,b)=\mathop{\min} \limits_{L\in \mathscr{L}} \mathop{\max} \limits_{z \in L}E(z)=\mathop{\max} \limits_{z \in L(a,b)} E(z)= E(s_1(a,b)).
\end{equation}

For complicated transition network graph of jamming PEL, we assert that the MEP connecting two minima can be derived from a MST of the graph. A proof is provided in the following two steps.

\subsubsection*{Lemma: The maximum-weight edge in a cycle is not contained in any MST.}  

\textbf{Proof:}  
Let \( G = (V, E) \) be an undirected connected graph with edge weights (energy) \( w: E \to \mathbb{R}_{\geq 0} \). Let \( C \subseteq E \) form a cycle, and let edge \( e_{\text{max}} = (u, v) \in C \) satisfy \( w(e_{\text{max}}) \geq w(e) \) for all \( e \in C \).  

Suppose for contradiction \( e_{\text{max}} \in T \), where \( T \) is an MST of \( G \). Removing \( e_{\text{max}} \) from \( T \) splits \( T \) into two disjoint connected components \( T_1 \) (containing \( u \)) and \( T_2 \) (containing \( v \)). Since \( C \) is a cycle, there exists \( e' = (x, y) \in C \setminus \{e_{\text{max}}\} \) with \( x \in T_1 \) and \( y \in T_2 \) (otherwise \( C \) cannot connect \( u \) and \( v \)). By definition of \( e_{\text{max}} \), \( w(e') < w(e_{\text{max}}) \).  

Define \( T' = (T \setminus \{e_{\text{max}}\}) \cup \{e'\} \). \( T' \) is connected (via \( e' \)), has \( |V| - 1 \) edges (hence acyclic), and \( w(T') = w(T) - w(e_{\text{max}}) + w(e') < w(T) \). This contradicts \( T \) being an MST, so \( e_{\text{max}} \notin T \).  

\subsubsection*{Theorem: The  $a$-$b$  path in an MST is an MEP connecting $a$ and $b$.}  

\textbf{Proof:}  
Let \( T \) be an MST of \( G \), and let \( L_{\text{MST}} \subseteq T \) be the unique $a$-$b$ path in \( T \). Let \( B = \max_{e \in L_{\text{MST}}} w(e) \), achieved by edge \( e_B \in L_{\text{MST}} \).  

Suppose for contradiction there exists an $a$-$b$ path \( L' \) with its maximum weight of edge \( B' < B \). The union \( L_{\text{MST}} \cup L' \) contains a cycle \( C \) that includes \( e_B \) (since \( L_{\text{MST}} \) and \( L' \) connect \( a \) and \( b \)). All edges in \( L' \) have weight \( \leq B' < B \), so \( e_B \) is the maximum-weight edge in \( C \).  

By the lemma, \( e_B \notin T \)—yet \( e_B \in L_{\text{MST}} \subseteq T \), and that is a contradiction. Consequently, no $a$-$b$ path has a maximum edge weight smaller than \( B \), so \( L_{\text{MST}} \) is indeed an MEP.

\section{Hierarchical clustering of local minima}

\begin{figure}[htb]
    \centering
    \includegraphics[scale=0.4]{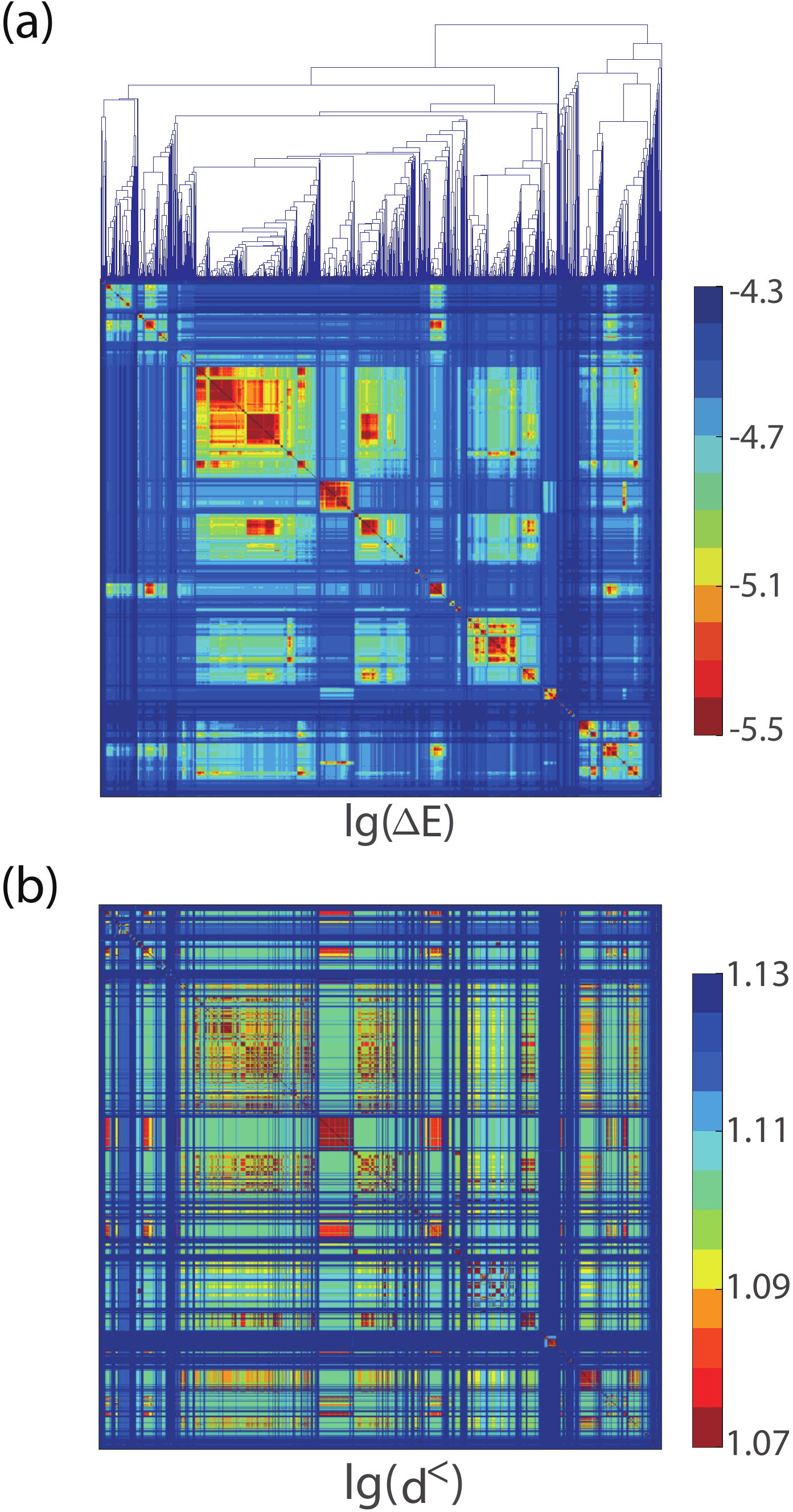}
    \caption{{\bf The system with the Hertzian interaction potential (volume fraction $\varphi=0.66$)}
 (a) $\Delta E$ matrix (associated with the barrier tree) and (b) $d^<$ matrix with the index of the local minima consistent with (a). The off-diagonal elements of both matrices are plotted in a logarithmic scale.
    }
    \label{smfig2}
\end{figure}

Fig.~\ref{smfig2} illustrates the energy barrier metric and $d^<$ metric of local minima in a large connected pathway map comprising 1229 minima. The minima are hierarchically clustered into a tree-structured dendrogram, constructed based on the two metrics using MATLAB's {\it dendrogram} function.

\clearpage

\section{Results of systems with the Harmonic interaction of different system sizes}

\begin{figure}[bth]
    \centering
    \includegraphics[width=\linewidth]{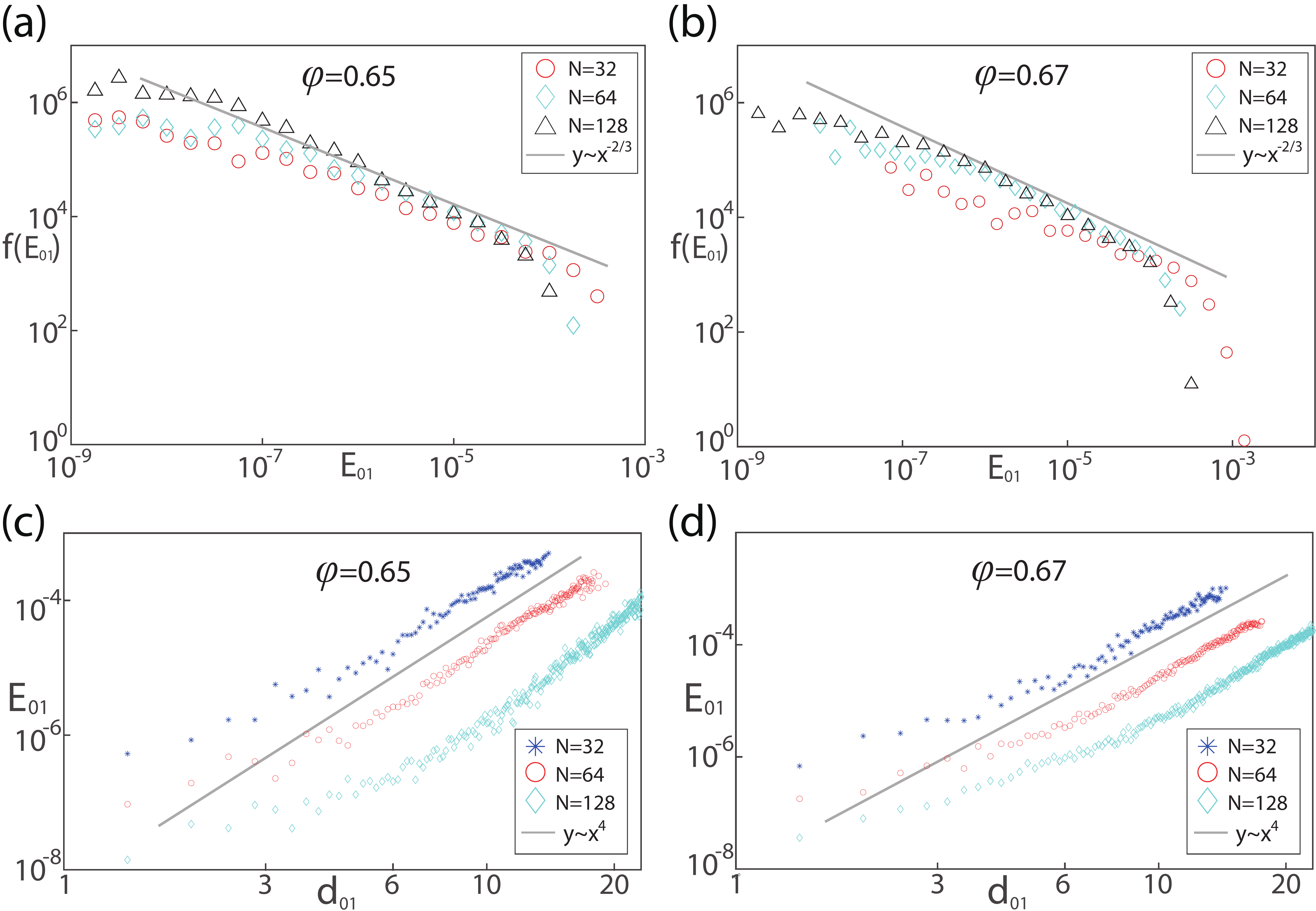}
    \caption{{\bf Systems of $N=32,64,128$ particles with the Harmonic interaction potential.} (a, b) The probability density function of $E_{01}$ is plotted for (a) $\varphi=0.65$  and (b) $\varphi=0.67$. (c, d) Scatter plot of the relation between $E_{01}$ and $d_{01}$ at (c) $\varphi=0.65$ and (d) $\varphi=0.67$.
    }
    \label{smfig3}
\end{figure}

Fig.~\ref{smfig3} shows additional results on the statistics of $E_{01}$.

\section{Derivation of the quartic relation $E_{01} \sim d_{01}^4$}

The derivation of the relation,
\begin{equation}
\label{eq:E_d}
E_{01} \sim d_{01}^4,
\end{equation}
consists of three basic steps. (i) The energy difference $E_{01}$ is proportional to the square of the length $L$, 
\begin{equation}
E_{01} \sim L^2,
\label{eq:E_L}
\end{equation}
where $L$ is 
the total length of a steepest gradient descent path between a 1-saddle and a minimum.  
(ii) The total path length $L$ is proportional to the total number of changed (breaking or closing) contacts $\delta n$,
\begin{equation}
L \sim \delta n.
\label{eq:L_n}
\end{equation}
(iii) According to the definition Eq.~(1), the distance $d$ is dominated by changed contacts, 
\begin{equation}
\label{eq:d01}
d_{01} \approx \sqrt{\sum_{b=1}^{\delta n}\vec{C}_b^2}\sim \delta n^{0.5}. 
\end{equation}
Combining (i-iii), we obtain Eq.~(\ref{eq:E_d}). Eq.~(\ref{eq:d01}) is a natural consequence of the definition of the distance metric. Next we examine Eqs.~(\ref{eq:E_L}) and~(\ref{eq:L_n})  numerically by solving the steepest gradient descent equation, Eq.~(\ref{eq:GD}), with a constant time step $10^{-2}$ (the elementary version of  SDA).

\begin{figure}[bth]
    \centering
    \includegraphics[width=0.9\linewidth]{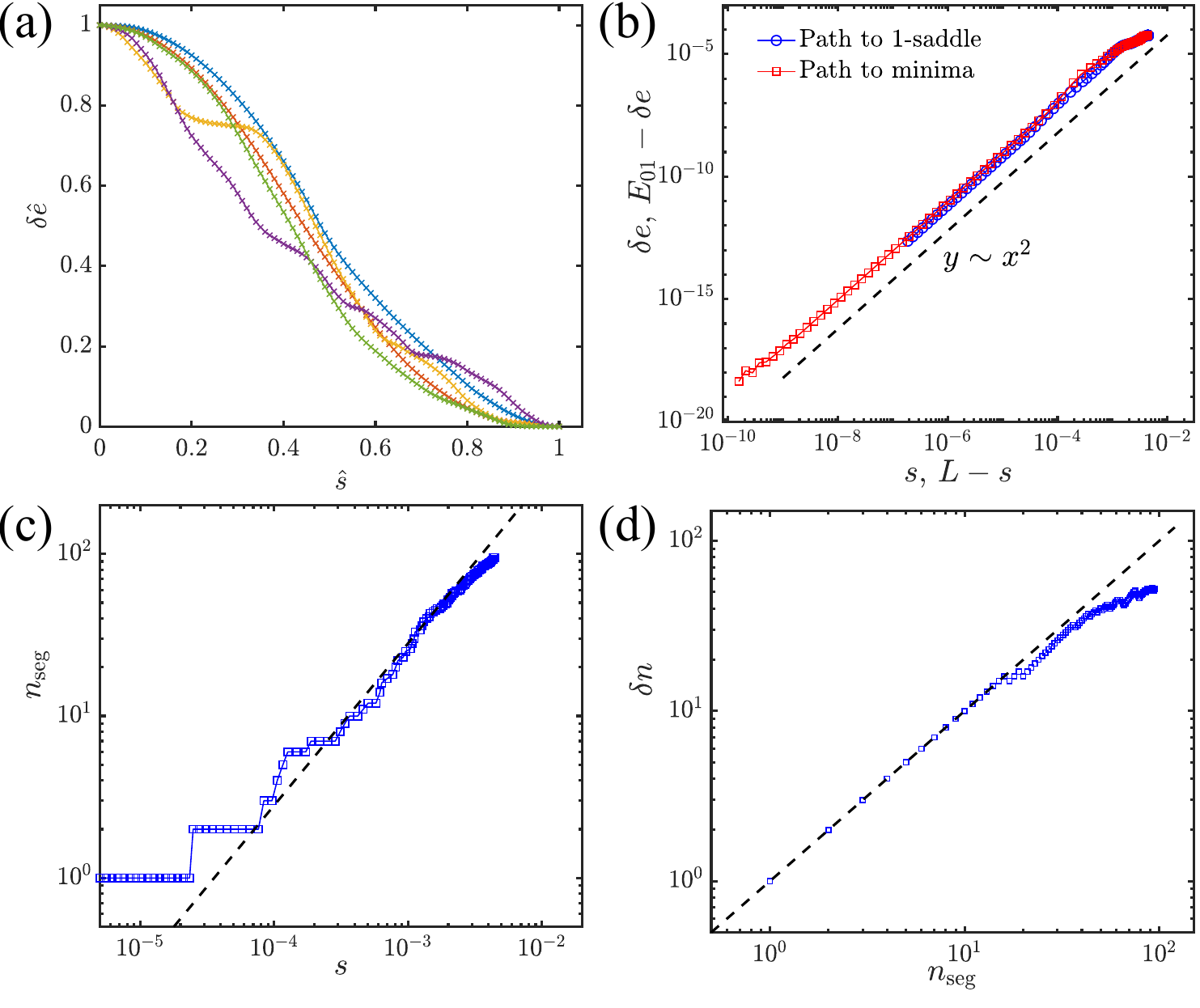}
    \caption{{\bf Evolution of the potential energy, path length, and contacts along steepest gradient descent paths.} 
    (a) The scaled reduction of potential energy along the scaled path length for several typical paths. 
    (b) The energy difference $\delta e$ (or $E_{01} - \delta e$) as a function of the path length $s$ (or $L - s$). The dashed line is the quadratic function.
    (c) The number of segments as a function of path length. The dashed line is the linear scaling relation, $n_{\rm seg} \sim s$.
    (d) The parameterized plot between $\delta n(s)$ and $n_{\rm seg}(s)$ for one typical path. The dashed line is $y = x$.
    All data are obtained for harmonic systems at $\varphi = 0.67$.
    }
    \label{sgd01path}
\end{figure}

In Fig.~\ref{sgd01path}(a), we show the scaled reduction of the potential energy, $\delta\hat{e} = \delta e/E_{01}$, where $0 \leq \delta e \leq E_{01}$, as a function of the scaled path length, $\hat{s} = s/L$, where $0<s<L$. The data are obtained from 
 several typical steepest gradient descent paths.
In Fig.~\ref{sgd01path}(b), we plot $\delta e$ as a function of $s$ (the path length to the 1-saddle), and $E_{01} - \delta e$ as a function of $L-s$ (the path length to the minimum), both of which are consistent with a quadratic relation, supporting Eq.~(\ref{eq:E_L}).

Next we examine Eq.~(\ref{eq:L_n}). As shown in Fig.~\ref{sgd01path}(a), each steepest gradient descent path consists of several segments. This is because the contact network changes during the energy minimization. Every time the contact network changes, the energy changes more rapidly. On the other hand, each segment corresponds to a stable contact network, during which the energy reduces slowly and smoothly. The accumulative number of segments is denoted by $n_{\rm seg}(s)$. 
A linear relationship, $n_{\rm seg} \sim s$ , is found in our simulations (see Fig.~\ref{sgd01path}c).
In Fig.~\ref{sgd01path}d, we plot $\delta n(s)$ as a function of $n_{\rm seg}(s)$, and find again a  linear relationship $\delta n \sim n_{\rm seg}$. 
The parameterized plot, $\delta n(n_{\rm seg})$, monotonically increases with contact  breaking.
The abrupt jumps in this curve is due to contact closing, which plays a minor role compared to the contact breaking.
Obviously, contact breaking dominates the increasing of $\delta n$ near the 1-saddle.
Combing these analyses, we have $L \sim n_{\rm seg} \sim \delta n$, verifying Eq.~(\ref{eq:L_n}).

\section{A phenomenological explanation of $f(E_{01}) \sim E_{01}^{-2/3}$}

\begin{figure}[bth]
    \centering
    \includegraphics[width=\linewidth]{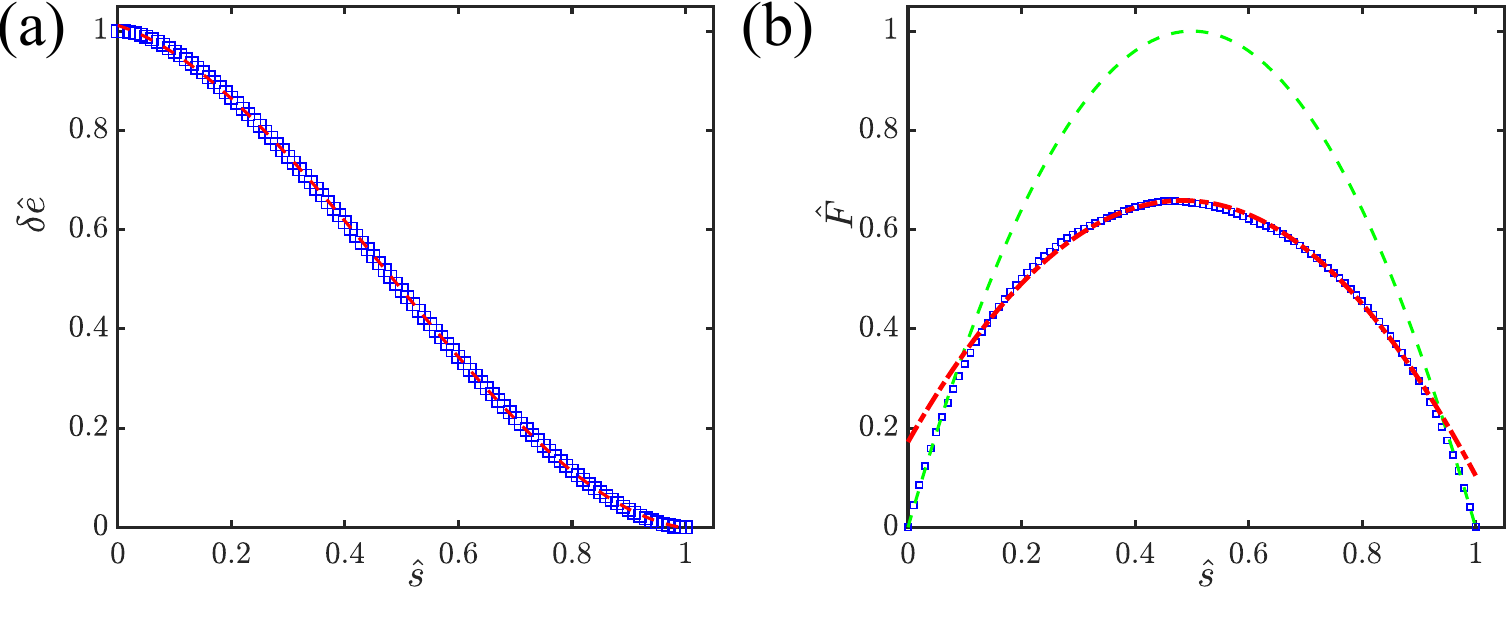}
    \caption{{\bf Evolution of the path-averaged energy and net force per particle along steepest gradient descent paths.} 
    (a) The path-averaged scaled reduction of energy. The dashed line is the third order polynomial fitting curve.
    (b) The path averaged scaled force $\hat{F}(\hat{s})$. The green dashed  line is the function, $y = 4x(1-x)$, and the red dash-dotted line is the quadratic fitting function, $y = -2.0 (s-0.48)^2+0.657$, for the interval $\hat{F}(\hat{s}) > 0.3$.
    All data are obtained for harmonic systems at $\varphi = 0.67$.
    }
    \label{avePath}
\end{figure}

As shown in Fig.~\ref{avePath}(a), the (averaged) energy function $\delta e(s)$ approximately follows a third order polynomial, 
\begin{equation}
 \delta e(s) = -\frac{1}{2} \lambda s^2 + \frac{1}{3} a s^3+ \ldots.
\label{eq:e}
\end{equation}
Further evidence is given by the force function. 
In Fig.~\ref{avePath}(b), we plot the scaled net force per particle $\hat{F}(\hat{s}) = |F(\hat{s})|/|F(\hat{s})|_{\rm max}$, averaged over paths at the same $\hat{s}$. 
The scaled force  can be well described by a quadratic function, $y = 4s(1-s)$, near the two ends, and by another quadratic function near the center.  The quadratic form of the force, which is the derivative of the energy, suggests that $\delta e(s)$ is at least third order, as in Eq.~(\ref{eq:e}). Solving for the stationary points of Eq.~(\ref{eq:e}), we have $E_{01} = \alpha \lambda^3/a^2$, where $\alpha = \frac{1}{6}$.

For individual samples, we assume that the energy function is still approximately a third order polynomial as Eq.~(\ref{eq:e}), but the coefficients fluctuate from sample to sample. Denoting their joint distribution by $f_{\lambda a}(\lambda,a)$, 
the distribution of $E_{01}$ can be then estimated by
\begin{equation}
f(E_{01}) \sim \int d\lambda da f_{\lambda a}(\lambda,a) \delta( \alpha \lambda^3/a^2 - E_{01}) \sim E_{01}^{-2/3} 
\end{equation}
where we have assumed that joint distribution can be factorized, $f_{\lambda a}(\lambda,a) \approx f_{\lambda}(\lambda) f_{a}(a) $,  and the distribution of $\lambda$ is not heavy-tailed.

\clearpage
\bibliographystyle{naturemag}
\bibliography{ref}